# DESIGN OF A 3 GHZ ACCELERATOR STRUCTURE FOR THE CLIC TEST FACILITY (CTF 3) DRIVE BEAM

G. Carron, E. Jensen, M. Luong[*)], A. Millich, E. Rugo, I. Syratchev, L. Thorndahl,
CERN, Geneva, Switzerland

*Abstract*

For the CLIC two-beam scheme, a high-current, long-pulse drive beam is required for RF power generation. Taking advantage of the 3 GHz klystrons available at the LEP injector once LEP stops, a 180 MeV electron accelerator is being constructed for a nominal beam current of 3.5 A and 1.5 µs pulse length. The high current requires highly effective suppression of dipolar wakes. Two concepts are investigated for the accelerating structure design: the "Tapered Damped Structure" developed for the CLIC main beam, and the "Slotted Iris – Constant Aperture" structure. Both use 4 SiC loads per cell for effective higher-order mode damping. A full-size prototype of the TDS structure has been built and tested successfully at full power. A first prototype of the SICA structure is being built.

## 1 INTRODUCTION

The power generation in the CLIC two-beam scheme, described in detail in [1], relies on drive beam accelerators (DBAs) operated at 937 MHz (30 GHz/32). The DBA for CTF 3 will operate at 3 GHz. This allows re-use of S-band equipment from the LEP injector which will be available once LEP stops.

To maximize RF generation efficiency, the drive beam accelerator will be operated at almost 100 % beam loading, i.e. over the length of each accelerating structure, the accelerating gradient will decrease to almost zero.

The CTF 3 drive beam accelerator will consist of approximately 20 accelerating structures, each of 32 cells and a total length of 1.3 m. It will operate in $2\pi/3$ mode and at a moderate accelerating gradient of 7 MV/m.

Two types of structure are being studied: the Tapered Damped Structure (TDS) has originally been designed for the CLIC main accelerator and was scaled down in frequency by a factor 10. This study is well advanced. This structure has, however, the disadvantage of its size (outer diameter 430 mm). Since some structures at the upstream end will have to fit into focusing solenoids, we are now also studying a slotted iris structure with an outer diameter of only 174 mm as an alternative. This approach is interesting also in view of the scaling to 937 MHz, and has the additional feature of a large constant iris aperture and consequently lower short-range transverse wake fields. We refer to the latter as SICA (Slotted Iris – Constant Aperture).

## 2 GENERAL DESCRIPTION

Both the TDS and the SICA structure are based on classical S-band cells. For the generic geometry, see Fig.1. Detuning and modulation of the group velocity (from 5 to 2.5 % over the length of the structure) are implemented by iris variation in the TDS (keeping the nose-cone size $x$ at zero), and by nose-cone variation in the SICA structure (keeping $a$ constant at 17 mm). In both cases, $b$ is adjusted for the correct phase advance of $\lambda/3$ at the operating frequency. The first cell has the same dimensions in both designs ($a$ = 17 mm, $x$ = 0).

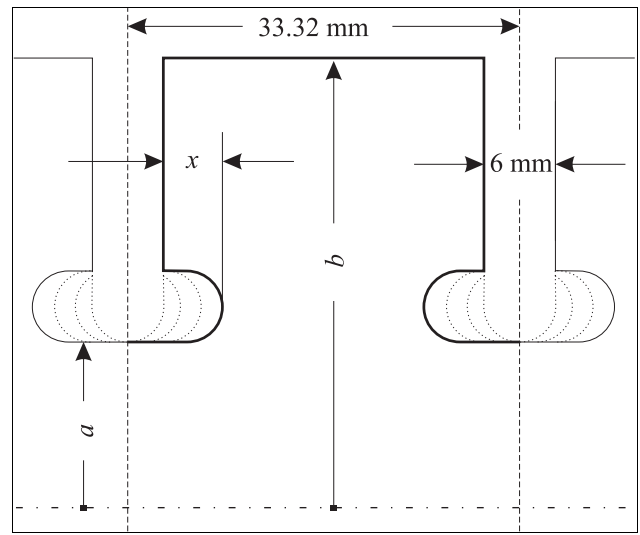

Figure 1: parameters of the cell geometry. The nose-cone size $x$ is zero for the TDS structure. Nose-cone sizes in different SICA cells are shown dotted.

The main difference between the two approaches is the coupling of the higher-order modes (HOMs) to the SiC loads. TDS uses wide openings in the outer cell wall, coupled to 4 waveguides with an axial E-plane, and with a cut-off above the operating frequency serving as high-pass filter for the HOMs. SICA on the other hand relies on geometrical mode-separation by 4 thin radial slots through the iris, coupling dipole modes to a ridged waveguide with its E-plane in the azimuthal direction. In

---
[*)] now at CEA-Saclay, DSM/DAPNIA/SEA, Gif-sur-Yvette, France

both cases, the SiC absorbers are wedge-shaped for good matching.

Both designs allow highly effective dipole mode damping. Fig. 2 shows the time domain MAFIA simulations of the transverse wake for a Gaussian bunch with a $\sigma$ of 2.5 mm, assuming the damping waveguides to be matched. These results were confirmed by frequency-domain calculations with HFSS, taking the properties of SiC into account: The first dipole mode of the TDS had a $Q$ of 18, that of the SICA structure had a $Q$ of 5.

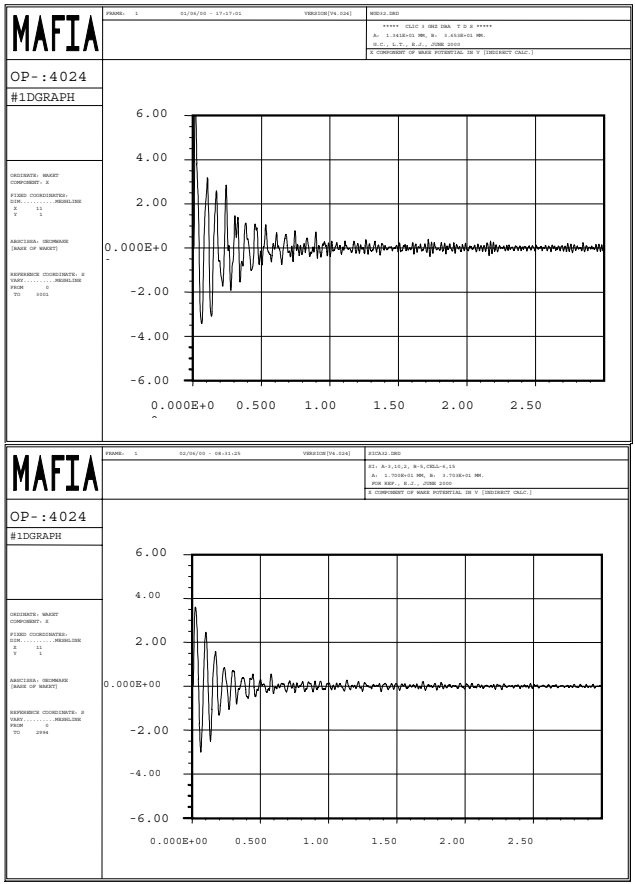

Figure 2: MAFIA time domain simulation of the dipole wake in the 32$^{nd}$ cell of TDS (top) and SICA structure (bottom). Abscissa in m behind bunch centre, ordinate in V/(2 pC)/m/mm. Note the smaller short-range wake field in the SICA structure.

## 3 TDS

The cells are supplemented with four 32 mm wide damping waveguides against transverse and longitudinal HOMs. The cell wall thickness is about 20 mm and the extruded copper waveguides are brazed into openings in the cell wall as shown in Fig. 3. The extruded waveguides constitute convenient housings at their outer extremities for the SiC absorbers that can be inserted through 16 mm mini-flanges after the final brazing of the structure. The absorbers will then either have been clamped or brazed onto metal holders. By introducing the SiC wedges after the final structure brazing, thermal strains on SiC bonds can be avoided and exchangeability is obtained.

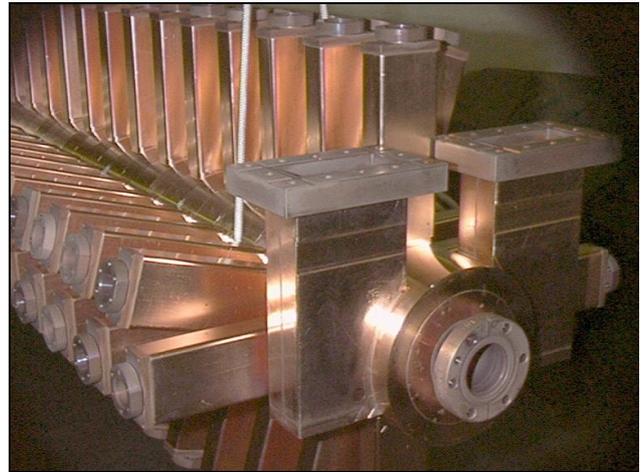

Figure 3: Brazed TDS ready for power testing

The prototype was brazed in 5 parts in a vertical position (2 couplers and the main body in 3 units). During that operation the damping waveguides with prebrazed end flanges were also bonded with the cells. Finally the 5 parts were brazed horizontally.

For future TDSs a single, uncomplicated vertical braze is foreseen at eutectic temperature, joining cells, damping waveguides and couplers. To avoid deformations (during the brazing) of the lowermost cells, the cell wall thickness will be increased to 35 mm, the total structure weight being ~160 kg.

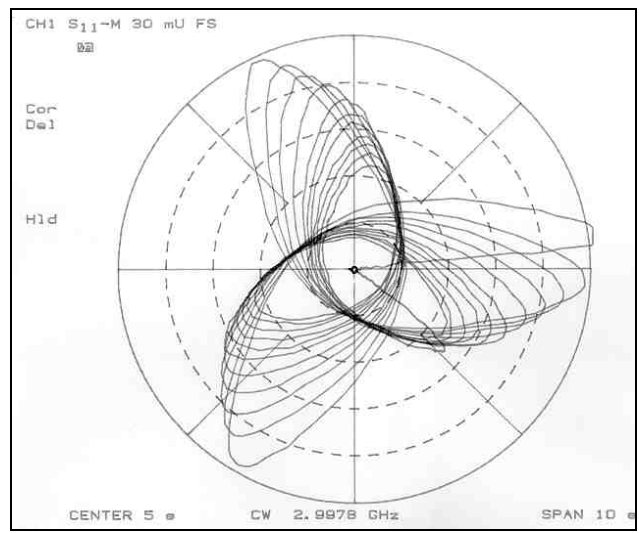

Figure 4: Bead-pull measurement: an error of about 40 ° between the 1$^{st}$ and last cell is measured in this reflection diagram, corresponding to a phase slip of ±10 ° between electron and wave over 32 cells. The loss in accelerating efficiency is less than 1 %. The cells are not equipped with dimple tuners.

### 3.1 Low-level measurements and power tests

Figs. 4 and 5 give measured low-level results for the brazed 32-cell TDS. Fig. 4 shows a measurable deviation from the ideal 240° phase advance per cell of the reflection coefficient, but even this error would lead to acceptable 1 % loss of accelerating efficiency. The matching of the input and output couplers over a bandwidth of 10 MHz is documented in Fig. 5.

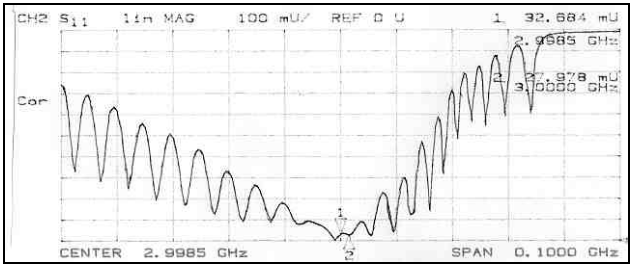

Figure 5: Reflection measurement: the input reflection loss is below –30 dB in a usable bandwidth of 10 MHz.

After brazing, the TDS was submitted to RF conditioning and reached the nominal power of 40 MW in less than 1 week, and subsequently 52 MW (the maximum power level available at CERN).

## 4 SICA

In the tapered damped structure, the waveguides between the accelerating cells and the SiC loads serve as high-pass filters, below cut-off for the accelerating mode, but transparent for higher-order modes, in particular the first dipole mode (at approximately 4.1 GHz). In order for this filter to work effectively, substantial waveguide length is required. This lead to an outer diameter of the 3 GHz TDS of approximately 430 mm.

As opposed to the "filter" type mode selection of the TDS, the SICA structure uses "geometric" mode selection [2]: a small radial slot in the iris does not intercept radial nor axial surface currents, so it will not perturb the accelerating $TM_{01}$ mode (nor any other $TM_{0n}$ mode). Dipole modes however have azimuthal current components which are intercepted and will thus induce a voltage across the slot. If this slot continues radially, cutting the outer cell wall, it can be considered as a waveguide, the cut-off of which can be made small by using a ridged waveguide.

Another concern was the short-range transverse wake, the strength of which is dominated by the iris aperture alone and hardly affected by detuning or damping. So we introduced nose-cones, varying in size from cell to cell to obtain the same linear variation of the group velocity (5 to 2.5 %) as in the conventionally detuned structure, keeping the iris aperture constant at Ø 34 mm. The overall detuning of the first dipole mode is slightly smaller than for the TDS, but sufficient. All other HOMs, longitudinal and transverse, show a larger detuning.

The ridged waveguides are machined into the cell, the slots by wire etching, the waveguide by milling. The SiC wedges are fixed with a specially designed clamp, avoiding thermal strains during brazing. Water cooling channels and dimple tuning are provided. The complete structure is brazed vertically and in one pass.

The SICA structure leads to a very compact design and its outer diameter of 174 mm allows the reuse of focusing solenoids from the LEP preinjector which have a bore of 180 mm.

A disadvantage of the SICA structure is an increased ratio of surface field to accelerating gradient, about 30 % higher than for the TDS structure at the downstream end. A further 20 % increase of the surface field will occur at the edges of the slots. Another critical issue is currently under study: since geometric mode selection is relying on symmetry, the demands on the symmetry of the structure are high. A quantitative sensitivity analysis is under way.

## ACKNOWLEDGEMENTS

The couplers are modified versions from the DESY S-Band collider structures. Thanks to DESY also for advice on brazing methods.

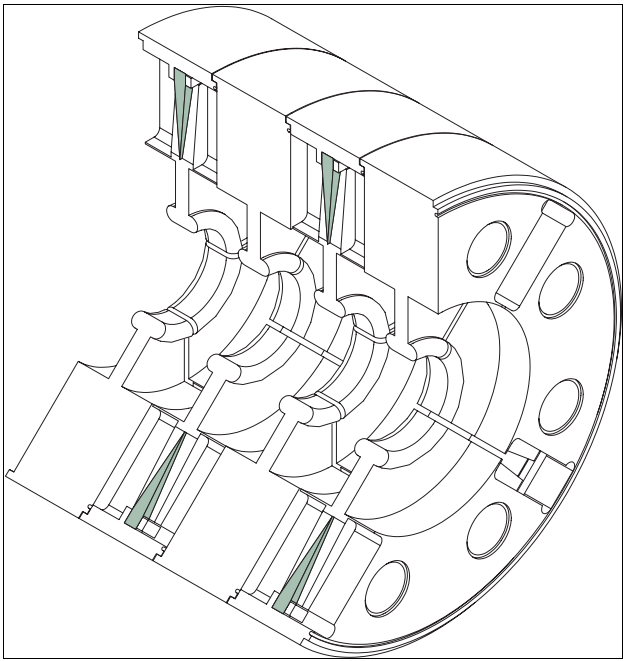

Figure 6: Cutaway view of 4 SICA cells. The outer diameter is 174 mm. The round holes are cooling water tubes; dimple tuning is provided (not shown).